\documentstyle[aps,preprint,floats,tighten,overcite]{revtex}
 
\input{psfig.tex}
\begin{document}

\title{The First Space-Based Gravitational-Wave Detectors}

\author{R.~R. Caldwell\footnote{Electronic address: \tt
caldwell@dept.physics.upenn.edu}}
\address{Department of Physics and Astronomy, University of Pennsylvania, \\
209 South 33rd Street, Philadelphia, PA 19104}
\author{Marc Kamionkowski\footnote{Electronic address: \tt
kamion@phys.columbia.edu} and
Leven Wadley\footnote{Electronic address: \tt leven@phys.columbia.edu}}
\address{Department of Physics, Columbia University, 538 West
120th Street, New York, NY 10027}

\maketitle

\begin{abstract}
Gravitational waves provide a laboratory for general relativity and a
window to energetic astrophysical phenomena invisible with electromagnetic
radiation. Several terrestrial detectors are currently under construction,
and a space-based interferometer is envisioned for launch early next
century to detect test-mass motions induced by waves of relatively short
wavelength. Very-long-wavelength gravitational waves can be detected using
the plasma in the early Universe as test masses;  the motion induced in
the plasma by a wave is imprinted onto the cosmic microwave background
(CMB). While the signature of gravitational waves on the CMB temperature
fluctuations is not unique, the {\it polarization} pattern can be used to
unambiguously detect gravitational radiation.  Thus, forthcoming CMB
polarization experiments, such as MAP and Planck, will be the first
space-based gravitational-wave detectors. 
\end{abstract}
  

\def\hatn{{\bf \hat n}}
\def\hatnprime{{\bf \hat n'}}
\def\hatnone{{\bf \hat n}_1}
\def\hatntwo{{\bf \hat n}_2}
\def\hatni{{\bf \hat n}_i}
\def\hatnj{{\bf \hat n}_j}
\def\vecx{{\bf x}}
\def\veck{{\bf k}}
\def\hatx{{\bf \hat x}}
\def\hatk{{\bf \hat k}}
\def\hatz{{\bf \hat z}}
\def\VEV#1{{\left\langle #1 \right\rangle}}
\def\cP{{\cal P}}
\def\noise{{\rm noise}}
\def\pix{{\rm pix}}
\def\map{{\rm map}}

One of the most spectacular predictions of general relativity is
the existence of gravitational waves. 
A gravitational wave conveys information about the motions
of mass and ripples in curvature---the shape of spacetime.
Detection of gravitational radiation would allow us to
probe ``invisible'' astrophysical phenomena hidden from view by
absorption of electromagnetic radiation.
Observations of the binary pulsar PSR1913+16, which confirm the
orbital-inspiral rate due to the emission of gravitational 
waves,\cite{binary} bring us tantalizingly close to this
goal.  However, we would still like to detect gravitational
radiation directly.  Thus, a variety of efforts are now under
way to detect gravitational waves.\cite{Thorne}  Here,
we show that forthcoming maps of the polarization of the cosmic
microwave background (CMB) can be used to detect very-long-wavelength
gravitational radiation.

Gravitational waves are detected by observing the motion they
induce in test masses.\cite{Bondi,MTW}  High-frequency ($1 -
10^4$ Hz) gravitational waves, produced by the inspiral and
catastrophic collision of astrophysical objects, may be
detectable by terrestrial laser interferometers (e.g.,
LIGO\cite{LIGO}) or resonant-mass antennae currently under
construction.  Low-frequency ($10^{-4} - 10^{-1}$ Hz)
gravitational waves, produced by the orbital motion of binaries,
could be detected by LISA,\cite{LISA} a space-based
interferometer targeted for launch {\it circa} 2015.

How can one detect ultra-low-frequency ($10^{-15} - 10^{-18}$
Hz) gravitational radiation, with
wavelengths comparable to the size of the
observable Universe?  The photon-baryon fluid in the early
Universe acts as a set of test masses for such waves.  A gravitational
wave in this frequency range alternately squeezes and
stretches the primordial
plasma.  Just as a resonant-mass detector is equipped with
electronics to monitor the oscillation modes of the test body,
the CMB photons are the electromagnetic signal upon which these
plasma motions are imprinted.

\begin{figure}[htbp]
\centerline{\psfig{figure=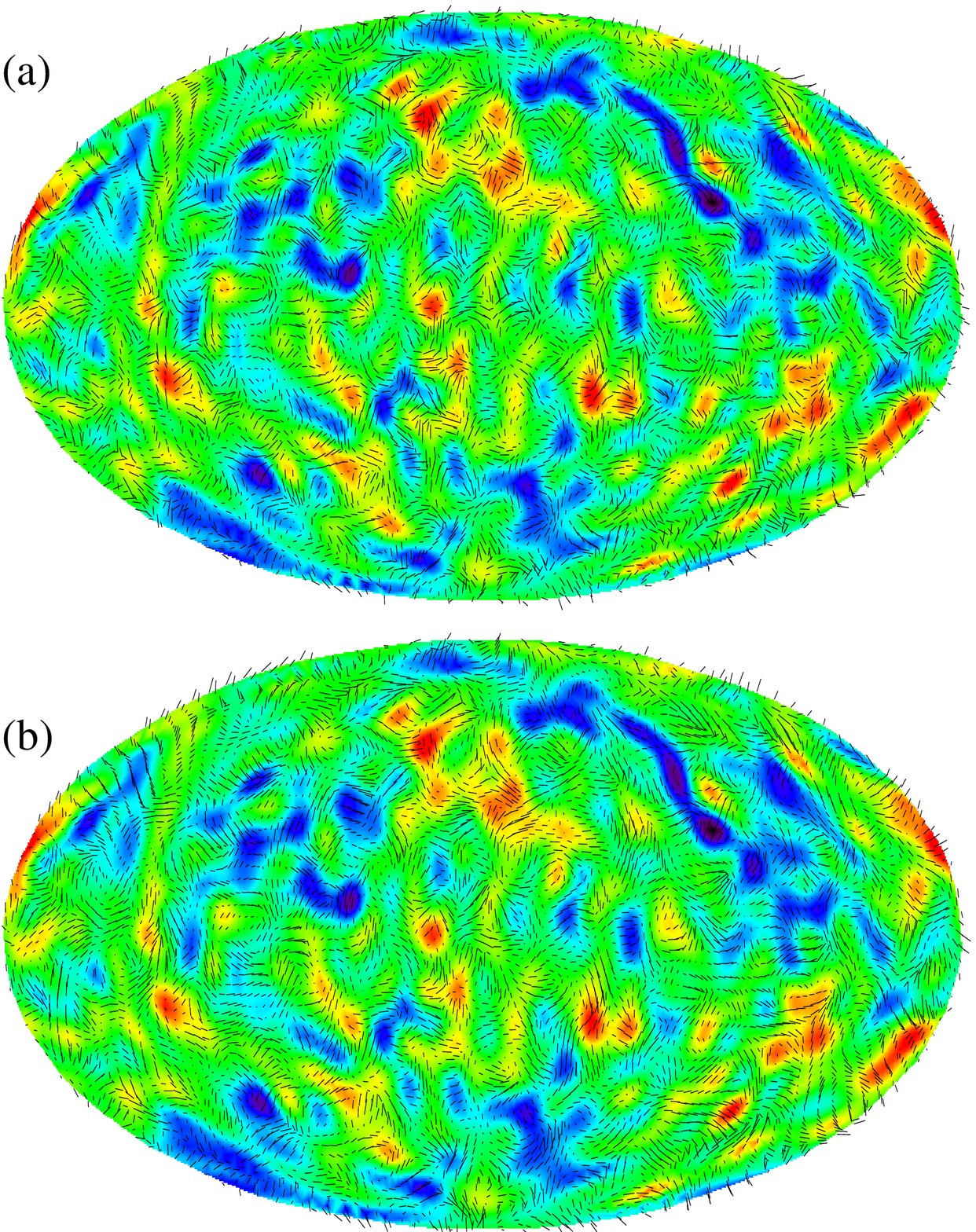,width=6in}}
\medskip
\caption{Simulated temperature-polarization maps of the CMB
	  sky for (a) density perturbations and (b)
	  long-wavelength gravitational waves.}
\label{fig:colormap}
\end{figure}

One might despair because cosmological density perturbations generate
equivalent fluctuations in the CMB temperature, nullifying the
possibility of using the temperature fluctuations to detect
gravitational waves (although it can be used to place
upper limits\cite{upperlimits}).  To illustrate, we show two simulated CMB
temperature and polarization maps in  Fig.~\ref{fig:colormap}.  The color
contrasts represent temperature fluctuations of roughly one part
in $10^5$ (red are hot spots and blue are
cold).  In (a), the temperature fluctuations are produced by a
spectrum of stochastic cosmological density perturbations. In
(b) we have found a plausible spectrum of stochastic
long-wavelength gravitational waves that produce precisely the
same temperature fluctuations.

Yet our hopes are restored upon realization that the motions in the
cosmological fluid generated by gravitational waves polarize the CMB in a
pattern that is distinct from that produced by density 
perturbations.\cite{letter,sz96b} Roughly speaking, the polarization ``vector''
field ${\vec P}(\hatn)$ (as a function of position $\hatn$ on the sky) can be
decomposed into a curl and curl-free part,
\begin{equation}
     {\vec P}(\hatn) = {\vec \nabla} A + \vec \nabla \times \vec
     B,
\label{vectorequation}
\end{equation}
where $A$ is a scalar function and ${\vec B}$ is a vector field. The curl and
curl-free parts of ${\vec P}(\hatn)$ can be isolated by taking the curl and
gradient of ${\vec P}$, respectively. Since density perturbations are scalar
perturbations to the spacetime metric, they have no handedness and   therefore
produce no curl in the CMB polarization field. Gravitational waves, however,
{\it do} have a handedness, so they {\it do} produce a curl.  By decomposing
the CMB polarization field into its curl and curl-free parts, one can
unambiguously detect gravitational waves. Again, to illustrate, 
the headless arrows in Fig.~\ref{fig:colormap} depict the
orientation and magnitude of the polarization at each point on
the sky.  We see that density perturbations and
gravitational waves that produce identical temperature
maps can be distinguished by the polarization pattern.

More precisely, the Stokes parameters $Q(\hatn)$ and $U(\hatn)$,
as a function of direction $\hatn=(\theta,\phi)$ on the sky, are
components of a symmetric trace-free (STF) $2\times2$ tensor,  
\begin{equation}
  {\cal P}_{ab}(\hatn)={1\over 2} \left( \begin{array}{cc}
   Q(\hatn) & -U(\hatn) \sin\theta \\
   \noalign{\vskip6pt}
   - U(\hatn)\sin\theta & -Q(\hatn)\sin^2\theta \\
   \end{array} \right),
\label{whatPis}
\end{equation}
that can be expanded
\begin{equation}
     {{\cal P}_{ab}(\hatn)\over T_0} = \sum_{\ell=2}^\infty\sum_{m=-\ell}^\ell
     \left[ a_{(\ell m)}^{\rm G}
     Y_{(\ell m)ab}^{\rm G}(\hatn) + a_{(\ell m)}^{\rm C}
     Y_{(\ell m)ab}^{\rm C} (\hatn) \right],
\label{Pexpansion}
\end{equation}
in terms of a basis $Y_{(\ell m)ab}^{\rm G}$, for the ``gradient'' (or
``curl-free''), and $Y_{(\ell m)ab}^{\rm C}$, for the ``curl,''
components of a STF
$2\times2$ tensor field, and $a_{(\ell m)}^{\rm G}$ and
$a_{(\ell m)}^{\rm C}$ are expansion coefficients.\cite{kkslongpaper}  

Consider a single gravitational wave with amplitude $h$ in the early Universe,
$+$ polarization, and comoving wavevector $\veck$ oriented in the $\hatz$
direction.  The curl component of the polarization pattern induced by this wave
will have expansion coefficients 
\begin{equation}
     a_{(\ell m)}^{{\rm C},+}(k,h) = - (2 \pi^2 i) \;h\;
     \sqrt{2\ell+1} \;
     (\delta_{m,2}- \delta_{m,-2})\; \Delta_{{\rm C}\ell}^{\rm
    (T)}(k,h=1).
\end{equation}
The functions $\Delta_{{\rm C}\ell}^{\rm (T)}(k,h=1)$ describe
the perturbations
to an isotropic photon distribution induced by a gravitational wave of
initially unit amplitude ($h=1$); they are obtained from equations for the
photon distribution in an expanding Universe with this gravitational wave.  The
precise form of $\Delta_{{\rm C}\ell}^{\rm (T)}(k,h=1)$ is only
weakly dependent on the cosmological model.  Only $m=\pm2$ modes
contribute since the effect of the gravitational wave is
symmetric under a $180^\circ$ rotation about the direction of
propagation.

\begin{figure}[htbp]
\centerline{\psfig{figure=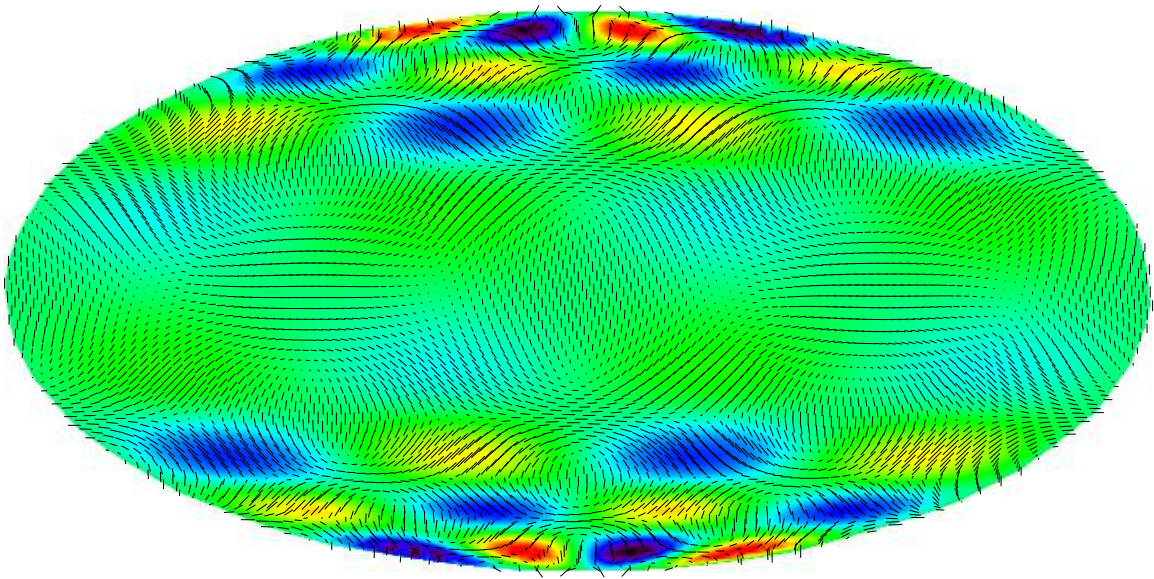,width=6in}}
\medskip
\caption{The temperature and curl component of the CMB polarization pattern
     produced by a single gravitational wave with wavenumber
     $k=4 H_0/c$, where $H_0$ is the Hubble constant,
     propagating in the $\hatz$ direction with $+$
     polarization. Such a CMB polarization pattern could {\it
     not} be mimicked by any combination of density perturbations.}
\label{fig:singlek}
\end{figure}

Fig.~\ref{fig:singlek} shows the
curl component of the polarization (as well as the temperature fluctuation)
produced by such a gravitational wave, a record of the
effect of the wave on the spherical surface of
last scatter.  The direction of propagation as well as the
polarization are clearly visible in this pattern; the wavelength
and the phase can also be inferred.  In this regard, the
CMB resembles a spherical resonant-mass detector\cite{spherical}
more than it resembles LISA or LIGO.

\begin{figure}[htbp]
\centerline{\psfig{figure=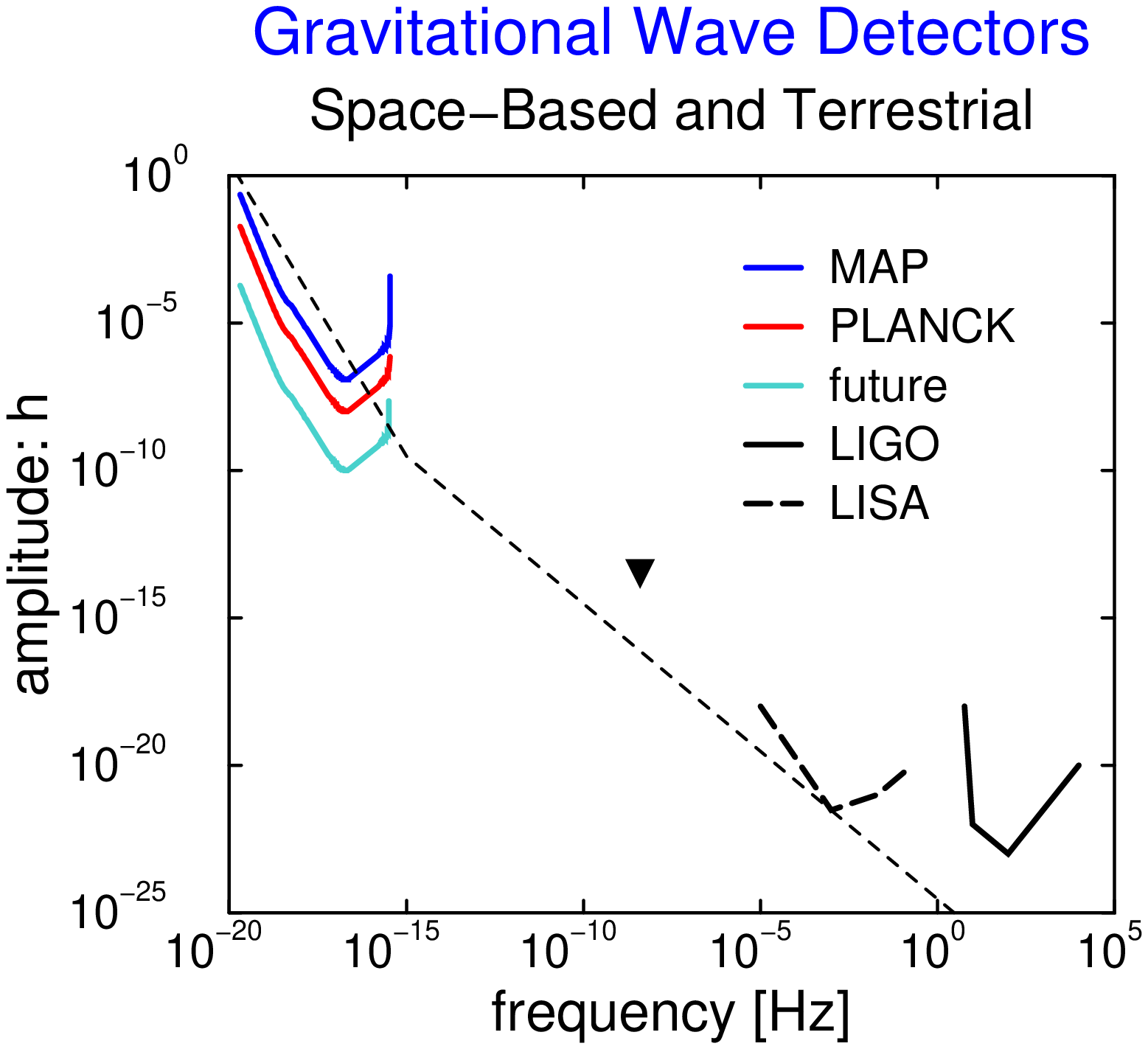,width=16.5cm}}
\caption{The smallest amplitude $h$ of a gravitational wave
     that can be detected at a frequency $f$ for LIGO, LISA, MAP, Planck, and a
     putative future CMB polarization experiment with 100 times the Planck
     sensitivity.  The short dashed line shows the largest
     scale-invariant stochastic gravitational-wave background
     consistent with COBE, and the triangle shows an upper limit
     {}from pulsar timing.}
\label{thefigure}
\end{figure}

What is the smallest dimensionless amplitude $h$ of a gravitational wave of
frequency $f$ that can be detected with such an experiment?  Suppose a CMB
polarization experiment measures $Q_i$ and $U_i$ at each of $i=1,2,...,N$ small
regions on the sky, each of area $4\pi/N$.  Then we have $2 N$ independent
measurements of $h$, one for each $Q_i$ and $U_i$, each with an
instrumental noise $\sigma$.  The minimum-variance estimator of
$h$ for the map is obtained from
the weighted average of all of these measurements, and the variance with which
$h$ can be determined is therefore $\sigma_h$, given by
\begin{equation}
     \left( {h \over \sigma_h} \right)^2 = \sum_i { \left[
     Q_i(h=1)\right]^2 + \left[U_i(h=1)\right]^2 \over
     \sigma^2} = 2 {N \over 4\pi \sigma^2}
     \sum_{\ell m}
     |a_{(\ell m)}^{\rm C}|^2.
\end{equation}
In Fig.~\ref{thefigure}, we plot the results for the
smallest amplitudes that can be detected at the $2\sigma$ level by 
the Microwave Anisotropy Probe (MAP),\cite{MAP} and the Planck 
Surveyor,\cite{PLANCK} CMB satellite experiments scheduled for launch,
respectively, by NASA in the year 2000 and by the European Space Agency five
years later.  The amplitude
detectable by a given CMB experiment is proportional to the
instrumental noise $\sigma$ in the detector.  If we extrapolate the rate
of progress in detector technology the past few
decades---roughly an order of magnitude per decade---into the
future, then it is plausible that the sensitivity could be
improved by a factor of 100 over that of Planck within the timescale for 
flight of LISA.  Thus, we also show in Fig.~\ref{thefigure} the
smallest amplitude $h$ that could be detected by such a future
CMB polarization experiment.  The smallest $h$
detectable by LIGO and LISA are also plotted. For reference, we
show the amplitude of the largest scale-invariant stochastic
gravitational-wave background consistent with
COBE\cite{upperlimits} (the short-dashed curve) and an upper
limit from pulsar timing\cite{pulsars} (the triangle).


Are there any promising sources of such long-wavelength gravitational
radiation?\cite{Allen}  And if so, what could we learn from them?  The most
encouraging and intriguing source is the spectrum of
gravitational waves produced from quantum fluctuations in the
spacetime metric (analogous to Hawking radiation) during
slow-roll inflation. If detected, these
waves would allow us to probe the early Universe well beyond the
epoch when it became opaque to electromagnetic radiation at
$z\simeq1100$, out to the inflationary epoch at redshifts $z\sim
10^{28}$, roughly $10^{-20}$ seconds after the big bang.  It can
also be shown that the amplitude of the stochastic
gravitational-wave background depends on the energy scale of
inflation.  If inflation has something to do with grand
unification, as many theorists surmise, then the amplitude
should be $h\sim10^{-5}$, possibly within reach of these CMB
experiments.
Other possible sources of ultra-low-frequency gravitational waves include
bubble collisions during a first-order phase transition in the early Universe
\cite{bubbles} or the action of topological defects \cite{topologicaldefects}. 
Thermal gravitational waves might be left over from the Planck era
\cite{Planckera}, and string theory-inspired alternatives/extensions to
inflation predict a unique spectrum of primordial gravitational radiation
\cite{nobigbang}.

The primary stated goals of MAP and Planck will be to determine
the geometry of the Universe and the origin of large-scale
structure.  But as we have shown here, these satellite
experiments, which will precede LISA by 10--20 years, may also
provide the first direct detection of gravitational
radiation.  The primordial plasma will provide the test masses
for this detector, and these gravitational-wave-induced motions
are isolated unambiguously in the CMB polarization.  Just as the
early Universe is the poor man's particle accelerator, the CMB
polarization will be the poor man's gravitational-wave detector.
If these experiments register a positive result, the
implications for general relativity, early-Universe cosmology,
and quantum theory in curved spacetime will be staggering.

\bigskip

This work was supported at Columbia University by
D.O.E. contract DEFG02-92-ER 40699, NASA ATP grant NAG5-3091,
and the Alfred P. Sloan Foundation, and at the University of
Pennsylvania by D.O.E. contract DEFG02-95-ER 40893.

\begin{references}

\bibitem{binary} J. H. Taylor, Rev. Mod. Phys. {\bf 66}, 711
     (1994). 

\bibitem{Thorne} E.g., K. Thorne, in {\it Particle and Nuclear
     Astrophysics and Cosmology in the Next Millennium},
     proceedings of the 1994 Snowmass Summer Study,
     eds. E. W. Kolb and R. D. Peccei (World Scientific,
     Singapore, 1995).

\bibitem{Bondi} H. Bondi, Nature {\bf 179}, 1072 (1957).

\bibitem{MTW} C. W. Misner, K. S. Thorne, and J. A. Wheeler,
     {\it Gravitation} (W. H Freeman \& Co, San Francisco,
     1973), pages 444--5.

\bibitem{LIGO} A. Abramovici et al., Science {\bf 256}, 325 (1992).

\bibitem{LISA} K. Danzmann {\it et al}, ``Laser Interferometer Space Antenna:
     Pre-Phase A Report'', February, 1996.

\bibitem{upperlimits} R. L. Davis et al., Phys. Rev. Lett. {\bf 69},
     1856 (1992); L. M. Krauss and M. White,
     Phys. Rev. Lett. {\bf 69}, 969 (1992).

\bibitem{letter} M.~Kamionkowski, A.~Kosowsky, and A.~Stebbins,
      Phys.\ Rev.\ Lett.\ {\bf 78}, 2058 (1997). 

\bibitem{sz96b} U.~Seljak and M.~Zaldarriaga, 
     Phys.\ Rev.\ Lett.\ {\bf 78}, 2054 (1997). 

\bibitem{kkslongpaper} M.~Kamionkowski, A.~Kosowsky, and
     A.~Stebbins, Phys. Rev. D {\bf 55}, 7368 (1997).

\bibitem{spherical} W. W. Johnson and S. M. Merkowitz,
     Phys. Rev. Lett. {\bf 70}, 2367 (1993).

\bibitem{MAP} {\tt http://map.gsfc.nasa.gov.}

\bibitem{PLANCK}
     {\tt http://astro.estec.esa.nl/SA-general/Projects/Planck/}

\bibitem{pulsars} V. M. Kaspi, J. H. Taylor, and M. F. Ryba,
     Astrophys. J. {\bf 428}, 713 (1994).
     
\bibitem{Allen} B. Allen, {\it The stochastic gravity-wave background:
     sources and detection}, in ``Proceedings of the Les Houches School
     on Astrophysical Sources of Gravitational Radiation,'' eds.
     J.-A. Marck and J.-P. Lasota (Cambridge University Press, 1997).

\bibitem{bubbles} E.g., C. Baccigalupi et al., Phys. Rev. D {\bf 56},
     4610 (1997).

\bibitem{topologicaldefects} R. R. Caldwell, R.A. Battye, and E.P.S. Shellard,
      Phys.Rev. D {\bf  54}, 7146 (1996);
     X. Martin and A. Vilenkin, Phys.Rev.Lett. {\bf 77} 2879, (1996).
      
\bibitem{Planckera} L. P. Grischuk, JETP {\bf 40}, 409 (1974).

\bibitem{nobigbang} G. Veneziano, hep-th/9802057.


\end{references}
\end{document}